\documentclass[english,aps,manuscript]{revtex4-1}
\usepackage[T1]{fontenc}
\usepackage[latin9]{inputenc}
\usepackage[active]{srcltx}
\usepackage{amstext}
\usepackage{amssymb}
\usepackage{esint}

\makeatletter
\@ifundefined{textcolor}{}
{%
 \definecolor{BLACK}{gray}{0}
 \definecolor{WHITE}{gray}{1}
 \definecolor{RED}{rgb}{1,0,0}
 \definecolor{GREEN}{rgb}{0,1,0}
 \definecolor{BLUE}{rgb}{0,0,1}
 \definecolor{CYAN}{cmyk}{1,0,0,0}
 \definecolor{MAGENTA}{cmyk}{0,1,0,0}
 \definecolor{YELLOW}{cmyk}{0,0,1,0}
 }

\makeatother

\usepackage{babel}
\begin{document}

\title{A Rotating Charged Black Hole Solution in $f\left(R\right)$ Gravity}

\author{Alexis Larrañaga}

\address{National Astronomical Observatory. National University of Colombia.}
\begin{abstract}
In the context of \emph{f(R) }theories of gravity, we address the
problem of finding a rotating charged black hole solution in the case
of constant curvature. The new metric is obtained by solving the field
equations and we show that the behavior of it is typical of a rotating
charged source. In addition, we analyze the thermodynamics of the
new black hole. The results ensures that the thermodynamical properties
in \emph{f(R)} gravities are qualitatively similar to those of standard
General Relativity. 


\end{abstract}
\maketitle

\section{Introduction}

An increasing attention has been paid recently to modified theories
of gravity in order to understand several open cosmological questions
such as the accelerated expansion of the universe and the dark matter
nature. A common option is to modify general relativity by adding
higher powers of the scalar curvature $R$, the Riemann and Ricci
tensors, or their derivatives \cite{fR1} in the lagrangian formulation.
Lovelock and $f\left(R\right)$ theories are some examples of these
attempts. Therefore, it is quite natural to ask about black hole existence
and its features in those gravitational theories. One can expect that
some signatures of black holes in these theories will be in disagreement
with the expected physical results of Einstein\textquoteright{}s gravity.
For those purposes, research on thermodynamical quantities of black
holes is of particular interest.

We will consider the $f\left(R\right)$ gravity theories in metric
formalism in Jordan\textquoteright{}s frame. The gravitational Lagrangian
is given by $R+f\left(R\right)$ where $f\left(R\right)$ is an arbitrary
function of the curvature scalar $R$. Einstein\textquoteright{}s
equations are usually fourth order in the metric \cite{fR2,fR3}.
Recently, the $f\left(R\right)$-Maxwell static black hole has been
obtained \cite{staticBH,staticBH1,staticBH2} and it has been shown
that all of its thermodynamic quantities are similar to those of the
Reissner-Nordström-AdS black hole when making appropriate replacements. 

In this paper, we find the rotating charged static black hole of $f\left(R\right)$
gravity in the case of constant curvature scalar and study some of
the thermodynamic aspects of the new solution.

\section{Rotating Charged Black Hole in $f\left(R\right)$ Theory}

Consider the action for $f(R)$ gravity with Maxwell term in four
dimensions

\begin{equation}
S=S_{g}+S_{M}
\end{equation}
where $S_{g}$ is the 4-dimensional gravitational action

\begin{equation}
S_{g}=\frac{1}{16\pi}\int\text{d}^{D}x\sqrt{\mid g\mid}(R+f(R))\label{S_g}
\end{equation}
and the electromagnetic action is

\begin{eqnarray}
S_{M} & = & -\frac{1}{16\pi}\int d^{4}x\sqrt{-g}\Big[F_{\mu\nu}F^{\mu\nu}\Big],\label{Actionfm}
\end{eqnarray}

where $g$ is the determinant of the metric, $R$ the scalar curvature
and $R+f(R)$ is the function defining the theory under consideration.
From the above action, the Maxwell equation takes the form 
\begin{equation}
\nabla_{\mu}F^{\mu\nu}=0.\label{maxw}
\end{equation}
while the field equations in the metric formalism are 

\begin{eqnarray}
R_{\mu\nu}\Big(1+f'(R)\Big)-\frac{1}{2}\Big(R+f(R)\Big)g_{\mu\nu}+\Big(g_{\mu\nu}\nabla^{2}-\nabla_{\mu}\nabla_{\nu}\Big)f'(R) & = & 2T_{\mu\nu}\label{equa}
\end{eqnarray}
where $R_{\mu\nu}$ is the Ricci tensor, $\nabla$ is the usual covariant
derivative and the stress-energy tensor of the electromagnetic field
is given by 
\begin{equation}
T_{\mu\nu}=F_{\mu\rho}F_{\nu}{}^{\rho}-\frac{g_{\mu\nu}}{4}F_{\rho\sigma}F^{\rho\sigma}
\end{equation}
with 
\begin{equation}
T^{\mu}~_{\mu}=0.
\end{equation}

Considering the constant curvature scalar $R=R_{0}$, the trace of
(\ref{equa}) leads to 
\begin{eqnarray}
R_{0}\Big(1+f'(R_{0})\Big)-2\Big(R_{0}+f(R_{0})\Big) & = & 0\label{eqR}
\end{eqnarray}
 which determines the negative constant curvature scalar as 
\begin{eqnarray}
R_{0} & = & \frac{2f(R_{0})}{f'(R_{0})-1}.\label{eqCR}
\end{eqnarray}

Using this relation in equation (\ref{equa}) gives the Ricci tensor

\begin{equation}
R_{\mu\nu}=\frac{1}{2}\left(\frac{f(R_{0})}{f'(R_{0})-1}\right)g_{\mu\nu}+\frac{2}{\Big(1+f'(R_{0})\Big)}T_{\mu\nu}.\label{eq:fieldequa}
\end{equation}

Inspired by the Kerr-Newman-AdS black hole solution, we introduce
the axisymmetric ansatz in Boyer-Lindquist- type coordinates $\left(t,r,\theta,\varphi\right)$,

\begin{equation}
ds^{2}=-\frac{B\left(r\right)}{\rho^{2}}\left[dt-\frac{a\sin^{2}\theta}{C\left(r\right)}d\varphi\right]^{2}+\frac{\rho^{2}}{B\left(r\right)}dr^{2}+\frac{\rho^{2}}{D\left(\theta\right)}d\theta^{2}+\frac{D\left(\theta\right)\sin^{2}\theta}{\rho^{2}}\left[adt-\frac{r^{2}+a^{2}}{C\left(r\right)}d\varphi\right]^{2}\label{eq:ansatz}
\end{equation}

with 

\begin{equation}
\rho^{2}=r^{2}+a^{2}\cos^{2}\theta
\end{equation}

and $B\left(r\right)$, \textbf{$C\left(r\right)$ }and \textbf{$D\left(\theta\right)$
}functions to be determined by the field equations. However, the field
equations alone are insufficient to determine all the unknown functions
uniquely. Since we are interested in solutions possesing a regular
horizon at $r=r_{+}$ we will impose the condition $B\left(r_{+}\right)=0$.
Additionally, we will consider that in the asymptotic region, the
metric will be flat if $R_{0}=0$ and $T_{\mu\nu}\rightarrow0$. 

Solving the Field equations (\ref{eq:fieldequa}) together with the
condition of constant curvature scalar, we obtain the solution 

\begin{eqnarray}
B & = & \Delta_{r}=\left(r^{2}+a^{2}\right)\left(1+\frac{R_{0}}{12}r^{2}\right)-2Mr+\frac{Q^{2}}{\left(1+f'\left(R_{0}\right)\right)}\\
C & = & \Xi=1-\frac{R_{0}}{12}a^{2}\\
D & = & \Delta_{\theta}=1-\frac{R_{0}}{12}a^{2}\cos^{2}\theta
\end{eqnarray}

and the line element can be written in the convenient form

\begin{equation}
ds^{2}=-\frac{\Delta_{r}}{\rho^{2}}\left[dt-\frac{a\sin^{2}\theta}{\Xi}d\varphi\right]^{2}+\frac{\rho^{2}}{\Delta_{r}}dr^{2}+\frac{\rho^{2}}{\Delta_{\theta}}d\theta^{2}+\frac{\Delta_{\theta}\sin^{2}\theta}{\rho^{2}}\left[adt-\frac{r^{2}+a^{2}}{\Xi}d\varphi\right]^{2}.\label{eq:rotcharmetric}
\end{equation}

Note that in the case $a=0$ the solution reproduces the charged rotating
black hole reported in \cite{staticBH,staticBH1,staticBH2} and it
is a generalization of the black hole for a limited case of $f\left(R\right)$
theories studied in \cite{rotatingfR}. The gauge field considered
has the potential

\begin{equation}
A_{t}(r)=-\frac{Qr}{\sqrt{\rho^{2}\Delta_{r}}}.
\end{equation}

\section{Thermodynamics}

To complete the analysis of the rotating charged solution, we will
calculate some thermodynamical quantities. The radius of the horizon
$r_{+}$ is defined by the condition $\Delta_{r}=0$, i.e.

\begin{equation}
\left(r_{+}^{2}+a^{2}\right)\left(1+\frac{R_{0}}{12}r_{+}^{2}\right)-2Mr_{+}+\frac{Q^{2}}{\left(1+f'\left(R_{0}\right)\right)}=0,\label{eq:horizondefinition}
\end{equation}

and it gives the horizon area 

\begin{equation}
A=\frac{4\pi\left(r_{+}^{2}+a^{2}\right)}{\Xi}=\frac{4\pi\left(r_{+}^{2}+a^{2}\right)}{1-\frac{R_{0}}{12}a^{2}}.\label{eq:area}
\end{equation}

The Hawking temperature is defined as

\begin{equation}
T=\frac{\kappa}{2\pi}
\end{equation}
with the surface gravity $\kappa$ given by

\begin{equation}
\kappa^{2}=-\frac{1}{2}\triangledown^{\mu}\chi^{\nu}\triangledown_{\mu}\chi_{\nu},
\end{equation}
where $\chi^{\nu}$ are null Killing vectors. The metric (\ref{eq:rotcharmetric})
has the Killing vectors $\xi^{\nu}=\partial_{t}$ and $\zeta^{\nu}=\partial_{\varphi}$
that are associated with the time translation and rotational invariance
respectively. Thus we define

\begin{equation}
\chi^{\nu}=\xi^{\nu}+\Omega\zeta^{\nu}
\end{equation}
 and we will find $\Omega$ imposing $\chi^{\nu}$ to be a null vector.
This gives

\begin{equation}
\chi^{\nu}\chi_{\nu}=g_{tt}+2\Omega g_{t\varphi}+\Omega^{2}g_{\varphi\varphi}=0,
\end{equation}
from which

\begin{equation}
\Omega=-\frac{g_{t\varphi}}{g_{\varphi\varphi}}\pm\sqrt{\left(\frac{g_{t\varphi}}{g_{\varphi\varphi}}\right)^{2}-\frac{g_{tt}}{g_{\varphi\varphi}}}
\end{equation}

and therefore at the event horizon $\Delta\left(r_{+}\right)=0$,
it reduces to

\begin{equation}
\Omega_{+}=\frac{a\Xi}{r_{+}^{2}+a^{2}}.\label{eq:omega}
\end{equation}
This gives the surface gravity

\begin{equation}
\kappa=\frac{1}{2\left(r_{+}^{2}+a^{2}\right)}\left.\frac{d\Delta_{r}}{dr}\right|_{r=r_{+}}
\end{equation}

\begin{equation}
\kappa=\frac{r_{+}^{2}-a^{2}-\frac{Q^{2}}{\left(1+f'\left(R_{0}\right)\right)}+\frac{R_{0}}{12}a^{2}r_{+}^{2}+\frac{R_{0}}{4}r_{+}^{4}}{2r_{+}\left(r_{+}^{2}+a^{2}\right)}
\end{equation}
and the corresponding Hawking temperature

\begin{equation}
T=\frac{r_{+}^{2}-a^{2}-\frac{Q^{2}}{\left(1+f'\left(R_{0}\right)\right)}+\frac{R_{0}}{12}a^{2}r_{+}^{2}+\frac{R_{0}}{4}r_{+}^{4}}{4\pi r_{+}\left(r_{+}^{2}+a^{2}\right)}.\label{eq:temperature}
\end{equation}

The same result is obtained by making an analytical continuation of
the Lorentzian metric by $t\rightarrow i\tau$ and $a\rightarrow ia$,
which gives the Euclidean section. Here, the regularity at $r=r_{+}$
requires the identification $\tau\sim\tau+\beta$ and $\varphi\sim\varphi+i\beta\Omega_{+}$
where 

\begin{equation}
\beta=\frac{4\pi r_{+}\left(r_{+}^{2}+a^{2}\right)}{r_{+}^{2}-a^{2}-\frac{Q^{2}}{\left(1+f'\left(R_{0}\right)\right)}+\frac{R_{0}}{12}a^{2}r_{+}^{2}+\frac{R_{0}}{4}r_{+}^{4}}=\frac{1}{T}.
\end{equation}

\subsection{Generalized Smarr Formula}

In order to obtain a generalized Smarr formula we find the toal energy
(mass) $E$ and angular momentum $J$ of the black hole by means of
Komar integrals. To do it, we use the Killing vectors $\frac{1}{\Xi}\frac{\partial}{\partial t}$
and $\frac{\partial}{\partial\varphi}$ to obtain

\begin{eqnarray}
E & = & \frac{M}{\Xi^{2}}\\
J & = & \frac{aM}{\Xi^{2}}.
\end{eqnarray}

Note that the time Killing vector has been normalized in order to
generate an $so\left(3,2\right)$ algebra from the corresponding conserved
quantities\@. Using this quantities and the horizon area (\ref{eq:area})
in equation (\ref{eq:horizondefinition}), we obtain 

\begin{equation}
E^{2}=\frac{A}{16\pi}+\frac{\pi}{A}\left[4J^{2}+\frac{Q^{4}}{\left(1+f'\left(R_{0}\right)\right)^{2}}\right]+\frac{Q^{2}}{2\left(1+f'\left(R_{0}\right)\right)}-\frac{R_{0}}{12}J^{2}-\frac{R_{0}A}{96\pi}\left[\frac{Q^{2}}{\left(1+f'\left(R_{0}\right)\right)}+\frac{A}{4\pi}-\frac{R_{0}A^{2}}{384\pi^{2}}\right].
\end{equation}

By identifying the entropy of the black hole as

\begin{equation}
S=\frac{A}{4}=\frac{\pi\left(r_{+}^{2}+a^{2}\right)}{1-\frac{R_{0}}{12}a^{2}}
\end{equation}

we obtain the generalized Smarr formula

\begin{equation}
E^{2}=\frac{S}{4\pi}+\frac{\pi}{4S}\left[4J^{2}+\frac{Q^{4}}{\left(1+f'\left(R_{0}\right)\right)^{2}}\right]+\frac{Q^{2}}{2\left(1+f'\left(R_{0}\right)\right)}-\frac{R_{0}}{12}J^{2}-\frac{R_{0}S}{24\pi}\left[\frac{Q^{2}}{\left(1+f'\left(R_{0}\right)\right)}+\frac{S}{\pi}-\frac{R_{0}S^{2}}{24\pi^{2}}\right].
\end{equation}

As is well known, this relation contains all the thermodynamical information
of the black hole. Therefore, one can define the quantities conjugate
to $S$, $J$ and $Q$ as the temperature, angular velocity and electric
potential respectively,

\begin{equation}
T=\left(\frac{\partial E}{\partial S}\right)_{J,Q}=\frac{1}{8\pi E}\left[1-\frac{\pi^{2}}{S^{2}}\left(4J^{2}+\frac{Q^{4}}{\left(1+f'\left(R_{0}\right)\right)^{2}}\right)-\frac{R_{0}}{6}\left(\frac{Q^{2}}{\left(1+f'\left(R_{0}\right)\right)}+\frac{2S}{\pi}-\frac{R_{0}S^{2}}{8\pi^{2}}\right)\right]\label{eq:T2}
\end{equation}

\begin{equation}
\Omega=\left(\frac{\partial E}{\partial J}\right)_{S,Q}=\frac{J}{E}\left(\pi S-\frac{R_{0}}{12}\right)\label{eq:omega2}
\end{equation}

\begin{equation}
\Phi=\left(\frac{\partial E}{\partial Q}\right)_{S,J}=\frac{Q}{2E\left(1+f'\left(R_{0}\right)\right)}\left[\frac{\pi}{S}\frac{Q^{2}}{\left(1+f'\left(R_{0}\right)\right)}+1-\frac{R_{0}S}{12\pi}\right].
\end{equation}

It is easy to verify that the relation (\ref{eq:T2}) for temperature
coincide with equation (\ref{eq:temperature}). However, it is not
true that the angular momentum (\ref{eq:omega2}) coincides with equation
(\ref{eq:omega}). In order to clarify this point, let us write the
metric (\ref{eq:rotcharmetric}) in the form

\begin{equation}
ds^{2}=-N^{2}dt^{2}+\frac{\rho^{2}}{\Delta_{r}}dr^{2}+\frac{\rho^{2}}{\Delta_{\theta}}d\theta^{2}+\frac{\vartheta^{2}\sin^{2}\theta}{\rho^{2}\Xi^{2}}\left(d\varphi-N^{\varphi}dt\right)^{2}
\end{equation}

where 

\begin{equation}
\vartheta^{2}=\left(r^{2}+a^{2}\right)\Delta_{\theta}-a^{2}\Delta_{r}\sin^{2}\theta
\end{equation}

\begin{equation}
N=\frac{\rho^{2}\Delta_{r}\Delta_{\theta}}{\vartheta^{2}}
\end{equation}

and

\begin{equation}
N^{\varphi}=\frac{a\Xi}{\vartheta^{2}}\left[\left(r^{2}+a^{2}\right)\Delta_{\theta}-\Delta_{r}\right].
\end{equation}

At the horizon, the function $N^{\varphi}$ coincides with $\Omega_{+}$but
asymptotically $\left(r\rightarrow\infty\right)$ it becomes
\begin{equation}
N^{\varphi}\rightarrow\frac{R_{0}}{12}a
\end{equation}
that can be interpreted as the angular velocity at infinity. Therefore,
the angular velocity defined in equation (\ref{eq:omega2}) actually
corresponds to the difference

\begin{equation}
\Omega=\Omega_{+}-\frac{R_{0}}{12}a
\end{equation}
as can easily be checked.

Finally, the thermal capacity $C$ at constant angular momentum and
charge is another thermodynamical quantity of interest. It is given
by

\begin{equation}
C=T\left(\frac{\partial S}{\partial T}\right)_{J,Q}=\frac{4\pi ETS}{1-4\pi T\left(2M+TS\right)-\frac{R_{0}}{6}\left(Q^{2}+\frac{3S}{\pi}-\frac{R_{0}S^{2}}{4\pi^{2}}\right)}.
\end{equation}

\section{Conclusion}

In this work we have obtained a rotating charged solution in $f\left(R\right)$
theory of gravity with constant curvature representing a black hole.
The new metric is obtained by solving the field equations and we have
also calculated some thermodynamical quantities. The behavior of the
new solution is typical of a rotating charged source and the analysis
show that the thermodynamical properties in \emph{$f(R)$} gravities
are qualitatively similar to those of standard General Relativity.
\\

\emph{Acknowledgements}

This work was supported by the National University of Colombia.

\end{document}